\documentclass[12pt]{article}
\usepackage{wider}
\usepackage{xcolor}
\usepackage{soul}
\usepackage{comment}
\usepackage{cite}
\usepackage{fnpct}

\begin{document}

\newcommand{\sqvb}{\ensuremath{ \langle \!\langle 0 |} }
\newcommand{\sqvk}{\ensuremath{ | 0 \rangle \!\rangle } }
\newcommand{\sqvn}{\ensuremath{ \langle \! \langle 0 |  0 \rangle \! \rangle} }
\newcommand{\wh}{\ensuremath{\widehat}}
\newcommand{\be}{\begin{equation}}
\newcommand{\ee}{\end{equation}}
\newcommand{\bea}{\begin{eqnarray}}
\newcommand{\eea}{\end{eqnarray}}
\newcommand{\ra}{\ensuremath{\rangle}}
\newcommand{\la}{\ensuremath{\langle}}
\newcommand{\rra}{\ensuremath{ \rangle \! \rangle }}
\newcommand{\lla}{\ensuremath{ \langle \! \langle }}
\newcommand{\str}{\rule[-.125cm]{0cm}{.5cm}}
\newcommand{\pr}{\ensuremath{^{\;\prime}}}
\newcommand{\ppr}{\ensuremath{^{\;\prime \prime}}}
\newcommand{\da}{\ensuremath{^\dag}}
\newcommand{\as}{^\ast}
\newcommand{\eps}{\ensuremath{\epsilon}}
\newcommand{\ve}{\ensuremath{\vec}}
\newcommand{\ka}{\kappa}
\newcommand{\non}{\ensuremath{\nonumber}}
\newcommand{\lf}{\ensuremath{\left}}
\newcommand{\rt}{\ensuremath{\right}}
\newcommand{\al}{\ensuremath{\alpha}}
\newcommand{\dfn}{\ensuremath{\equiv}}
\newcommand{\ga}{\ensuremath{\gamma}}
\newcommand{\ti}{\ensuremath{\tilde}}
\newcommand{\wti}{\ensuremath{\widetilde}}
\newcommand{\hs}{\ensuremath{\hspace*{.5cm}}}
\newcommand{\bet}{\ensuremath{\beta}}
\newcommand{\om}{\ensuremath{\omega}}
\newcommand{\kp}{\ensuremath{\kappa}}

\newcommand{\cO}{\ensuremath{{\cal O}}}
\newcommand{\cS}{\ensuremath{{\cal S}}}
\newcommand{\cF}{\ensuremath{{\cal F}}}
\newcommand{\cX}{\ensuremath{{\cal X}}}
\newcommand{\cZ}{\ensuremath{{\cal Z}}}
\newcommand{\cG}{\ensuremath{{\cal G}}}
\newcommand{\cR}{\ensuremath{{\cal R}}}
\newcommand{\cV}{\ensuremath{{\cal V}}}
\newcommand{\cC}{\ensuremath{{\cal C}}}
\newcommand{\cP}{\ensuremath{{\cal P}}}
\newcommand{\cH}{\ensuremath{{\cal H}}}
\newcommand{\cN}{\ensuremath{{\cal N}}}
\newcommand{\cE}{\ensuremath{{\cal E}}}

\newcommand{\pup}{\ensuremath{^{(p)}}}
\newcommand{\prpr}{\ensuremath{\prime \prime }}

\newcommand{\hsp}{\ensuremath{\hspace*{5mm} }}
\newcommand{\sbp}{\ensuremath{_{[p]} }}

\newcommand{\erf}{\mathop{\mathrm{erf}}}

\newcommand{\xxx}[1]{}
\newcommand{\yyy}[1]{}
\newcommand{\zzz}[1]{}
\newcommand{\xyz}[1]{}

\title{\bf Reverse quantum speed limit and minimum Hilbert space norm}

\author{
Mark A. Rubin\\
\mbox{}\\
markallenrubin@yahoo.com\\
}
\date{\mbox{}}
\maketitle
\begin{abstract}
The reverse quantum speed limit (RQSL)  gives an upper limit to the time for evolution between initial and final quantum states.  We show that, in conjunction with the existence of a minimum time scale, the RQSL implies a lower limit to the norm of the change in a quantum state, and confirm that this limit is satisfied in two-state and ideal-measurement models.  Such a lower limit is of  relevance for interpretational issues in probability and for understanding the meaning of probability in Everett quantum theory.


\mbox{}


\end{abstract}

\section{Introduction}\label{SecIntro}
\xxx{SecIntro}

The spacetime that provides the mathematical framework within which both classical and quantum fields are defined is a continuum. Nevertheless, it has been argued for decades that  the combination of quantum mechanics and general relativity implies the existence of a minimum length scale at or near the  Planck
 length.\footnote{A  device-independent argument for such a minimum length scale is presented in
  \cite{Calmetetal04}.
  For a review of this subject, see \cite{Hossenfelder13}.}
Relativistic covariance then implies the existence of a  corresponding minimum time scale comparable in magnitude to the Planck time \cite{BHZ05, FaizalKhalilDas16}.

Buniy, Hsu and Zee \cite{BHZ05}  have argued that the existence of a minimum length scale implies a minimum scale to the norms of states in quantum-mechanical Hilbert space,  such that components of the wavefunction with norm below this scale can be removed from the wavefunction.\footnote{Recently Calmet and Hsu \cite{CalmetHsu21} have presented an argument for such a Hilbert-space minimum norm based on a minimum scale to angular measurement coming from quantum mechanics and general relativity.}  The existence of such a minimum Hilbert-space norm 
provides a  solution to the problem of ``maverick'' states\cite{DeWitt70} in Everett quantum theory\cite{Everett57, DeWittGraham73, Price95, Barrett99, Vaidman02,  HewittHorsman09, Saundersetal10, Wallace12, Wallace13} and thus allows for an understanding of the origin and nature of probability in that theory  \cite{BHZ06, Hsu12, Hsu17, Hsu21, Rubin21}.

%

Hsu\cite{Hsu21} 
has presented an estimate for the norm of the change in a quantum state in the presence of a minimum time scale. In the present note we employ the reverse quantum speed limit (RQSL) of Mohan and Pati \cite{MohanPati20} to derive a lower limit to this norm.

\section{Reverse quantum speed limit}\label{SecRQSL}
\xxx{SecRQSL}

The RQSL gives an upper limit to the time $T$\/ for an initial state $|\psi(0)\ra$\/ to evolve to a final state $|\psi(T)\ra$\/.  
The expression for this limit involves the reference section $|\chi(t)\ra$\/, a function of the state vector $|\psi(t)\ra$\/ and the initial time (here taken to be $t=0$\/) defined as
\be
|\chi(t)\ra = \frac{\la \psi(t) | \psi(0) \ra }{| \la \psi(t) | \psi(0) \ra |} |\psi(t)\ra. \label{refsecdef}
\ee
\xxx{refsecdef}
\yyy{DHS-6, last box}
The length of the reference section for evolution from time $t=0$\/ to time $t=T$\/ is defined to be
\be
l(\chi(t))|_{0}^{T} =\int_{0}^{T} \la\dot{\chi}(t)|\dot{\chi}(t)\ra^{\frac{1}{2}} dt. \label{refseclendef}
\ee
\xxx{refseclendef}
\yyy{p. DHS-6, box 2}
In terms of the length (\ref{refseclendef}), the RQSL  is
\be 
T \leq \frac{\hbar l(\chi(t))|_{0}^{T} }{\Delta H},\label{RQSL}
\ee
\xxx{RQSL}
\yyy{p. DHS-6, box 1}
where $\Delta H$\/ is the square root of the expectation value of the variance of the Hamiltonian $\wh{H}$\/ in the state $|\psi(t)\ra$\/:
\be
\Delta H=\left(\la \psi(t)|\wh{H}^2|\psi(t)\ra-\la \psi(t)|\wh{H}|\psi(t)\ra^2\right)^\frac{1}{2}.\label{DeltaHdef}
\ee
\xxx{DeltaHdef}
\yyy{p. DHS-16, box 2}
Here we will only consider the case of time-independent Hamiltonians,\footnote{For the generalization to time-dependent Hamiltonians, see 
\cite{MohanPati20}.}
in which case $\Delta H$\/ is time-independent and the formula (\ref{DeltaHdef}) can be evaluated at any convenient time, in particular at $t=0$\/.

\section{Minimum Hilbert space norm}

Consider the time evolution to occur over a time interval $0 \leq t \leq \Delta t$\/ short enough so that we can approximate the derivative of the reference section (\ref{refsecdef}) during this interval as 
\be
|\dot{\chi}(t)\ra = \frac{|\chi(\Delta t)\ra -|\chi(0)\ra }{\Delta t}. \label{chiapprox}
\ee
\xxx{chiapprox}
\yyy{p. DHS-6, box 3}
Clearly the range of applicability of subsequent results will depend on the range of validity of the approximation (\ref{chiapprox}); we will return to this issue below.  

Using (\ref{chiapprox}) in (\ref{refseclendef}) with $T=\Delta t$\/,
\bea
l(\chi(t))|_{0}^{\Delta t}&=&\left[\left(\frac{\la\chi(\Delta t)| -\la\chi(0)| }{\Delta t}\right)\left(\frac{|\chi(\Delta t)\ra -|\chi(0)\ra }{\Delta t}\right)\right]^\frac{1}{2}\Delta t \nonumber \\
&=&\left[\left(\la\chi(\Delta t)| -\la\chi(0)| \right)\left(|\chi(\Delta t)\ra -|\chi(0)\ra \right)\right]^\frac{1}{2} \label{refseclen2}
\eea
\xxx{refseclen2}
\yyy{p. DHS-6, box 5}
Using (\ref{refsecdef}) in (\ref{refseclen2}), along with the normalization of $|\psi(t)\ra$\/,  yields  
\be
l(\chi(t))|_{0}^{\Delta t} =  \left[\la\psi(\Delta t)|\psi(\Delta t)\ra  -2|\la\psi(\Delta t)|\psi(0)\ra| +1\right]^\frac{1}{2}\label{refseclen3}
\ee
\xxx{refseclen3}
\yyy{p. DHS-8}
Expressing $|\psi(\Delta t )\ra$\/ in terms of $|\Delta \psi (\Delta t)\ra$\/, the change in $|\psi(t)\ra$\/, i.e. 
\be
|\psi(\Delta  t)\ra=|\psi(0)\ra + |\Delta \psi (\Delta t) \ra, \label{psitodeltapsi}
\ee
\xxx{psitodeltapsi}
\yyy{p. DHS-9, box 2}
and defining
\be
z=1+\la\Delta \psi (\Delta t) |\psi(0)\ra \label{zdef}
\ee
\xxx{zdef}
\yyy{p. DHS-12, box 4, p. DHS-31 box 3}
we obtain from (\ref{refseclen3})
\be
\left(l(\chi(t))|_{0}^{\Delta t}\right)^2 = \la \Delta \psi (\Delta t)| \Delta \psi (\Delta t)\ra + 2\left({\rm Re}(z) -|z|\right). \label{refseclenitoz}
\ee
\xxx{refseclenitoz}
\yyy{p. DHS-31, last box}
From the RQSL, eq. (\ref{RQSL}),
\be
\left(l(\chi(t))|_{0}^{\Delta t}\right)^2 \ge \left(\frac{\Delta t \Delta H}{ \hbar}\right)^2,\label{RSQL2}
\ee
\xxx{RSQL2}
\yyy{p. DHS-31, next-to-last box}
so, with (\ref{refseclenitoz}), we have
\be
\la\Delta \psi (\Delta t) | \Delta \psi (\Delta t) \ra -\left(\frac{\Delta t \Delta H}{ \hbar}\right)^2 \ge 2\left(|z|-{\rm Re}(z)\right). \label{almostthere}
\ee
\xxx{almostthere}
\yyy{p. DHS-31, box 1}
But for any complex number $z$\/
\be
|z|-{\rm Re}(z) \ge 0, \label{complexnumbers}
\ee
\xxx{complexnumbers}
\yyy{p. DHS-33}
so 
\be
\la \Delta \psi (\Delta t)| \Delta \psi(\Delta t) \ra^\frac{1}{2} \ge \frac{\Delta t \Delta H}{\hbar}. \label{normlimit}
\ee
\xxx{normlimit}
\yyy{ p. DHS-12, last box}

Now suppose that time is in fact discrete; i.e., there is a lower limit $\Delta t_{min}$\/ during which time evolution can take place. Then
\be
\la \Delta \psi (\Delta t_{min})| \Delta \psi (\Delta t_{min})\ra^\frac{1}{2} \ge {\rm NormLim_{RQSL}},
\label{normlimit2}
\ee
\xxx{normlimit2}
where
\be
{\rm NormLim_{RQSL}} = \frac{\Delta t_{min} \Delta H}{\hbar}. \label{NLRQSLdef}
\ee
\xxx{NLRQSLdef}
That is, 
there is no Hamiltonian with expected variance $\left(\Delta H\right)^2$\/ in  state $|\psi(0)\ra$\/ which will in a time $\Delta t_{min}$\/ evolve $|\psi(0)\ra$\/ into a state 
$|\psi(\Delta t_{min})\ra$\/ closer in norm to $|\psi(0)\ra$\/ than ${\rm NormLim_{RQSL}}$\/.\footnote{Note that the estimate given by Hsu \cite[eq. (6)]{Hsu21} for the norm of the change in the state, which in our notation is  $\Delta t_{min} \la\psi(0)|\wh{H}^2|\psi(0)\ra^\frac{1}{2}/\hbar$\/, will always be greater than or equal to   ${\rm NormLim_{RQSL}}$\/.  Hsu argues that by considering the estimates for all possible initial states in a finite-dimensional Hilbert space one may arrive at a lower bound on the norm.  The minimum norm given in the present paper applies to  all states with a specified value of $\Delta H$\/.}

\section{Examples}

Since time appears to us to be continuous rather than discrete, we will take the approximation (\ref{chiapprox}) and the conclusion (\ref{normlimit2}) that follows from it be valid provided $\Delta t_{min}$\/ is much smaller than the smallest time scale characterizing changes in  $|\psi(t)\ra$\/.  We consider two examples:

\subsection{Two-state system}\label{SubsecTwoState}
\xxx{SubsecTwoState}

The Hilbert space $\cH_2$\/ is spanned by two vectors, $|\cS_1\ra$\/ and  $|\cS_2\ra$\/, which we take to be normalized eigenstates of the Hamiltonian $\wh{H}_{\cH_2}$\/:
\be
\wh{H}_{\cH_2}|\cS_i\ra=E_i|\cS_i\ra,\hspace*{1cm} \la \cS_i|\cS_j\ra=\delta_{i,j},\hspace*{1cm}i,j=1,2.\label{Hilbertspace2}
\ee
\xxx{Hilbertspace2}
\yyy{p. DHS-15, boxes 2,4,5}
The normalized initial state vector is
\be
|\psi_{\cH_2}(0)\ra=c_1|\cS_1\ra+c_2|\cS_2\ra,\hspace*{1cm}|c_1|^2+|c_2|^2=1.\label{psi0HS2}
\ee
\xxx{psi0HS2}
\yyy{p. DHS-15, box 1}
The lower limit to the norm of the change in the state vector from $t=0$\/ to $t=\Delta t_{min}$\/ is, from (\ref{DeltaHdef}), (\ref{NLRQSLdef}), (\ref{Hilbertspace2}) and (\ref{psi0HS2}),
\be
{\rm NormLim}_{{\rm RQSL} - \cH_2}=\left[|c_1|^2E_1^2 + |c_2|^2E_2^2-\left(|c_1|^2E_1+|c_2|^2E_2\right)^2\right]^\frac{1}{2}\frac{\Delta t_{min}}{\hbar}. 
\label{NLHS2}
\ee
\xxx{NLHS2}
\yyy{p. DHS-17, last box}


The exact state vector for $t \ge 0 $\/ is, from (\ref{Hilbertspace2}), (\ref{psi0HS2}) and the Schr\"{o}dinger equation,
\be
|\psi_{\cH_2}(t)\ra=c_1\exp\left(\frac{-iE_1 t}{\hbar}\right)|\cS_1\ra + c_1\exp\left(\frac{-iE_2 t}{\hbar}\right)|\cS_2\ra.\label{psitHS2}
\ee
\xxx{psitHS2}
\yyy{p. DHS-15, box 6}
So, the shortest time scale characterizing changes in $|\psi_{\cH_2}(t)\ra$\/ is
\be
\Delta t_{\cH_2}=2\pi \hbar/\max_i(|E_i|),\label{DeltatHS2}
\ee
\xxx{DeltatHS2}
\yyy{p. DHS-18, boxes 2-6}
in that $|\psi(t)\ra_{\cH_2}$\/ will be approximately constant over time intervals much smaller than (\ref{DeltatHS2}).

Using (\ref{psitodeltapsi}), (\ref{psi0HS2}) and (\ref{psitHS2}),
\be
\la\Delta \psi_{\cH_2}(\Delta t)|\Delta \psi_{\cH_2}(\Delta t)\ra^\frac{1}{2}=
          \left[2-2|c_1|^2\cos\left(\frac{E_1\Delta t}{\hbar}\right)-2|c_2|^2\cos\left(\frac{E_2\Delta t}{\hbar}\right)\right]^\frac{1}{2}.
\ee
If 
\be 
\Delta t \ll \Delta t_{\cH_2} \label{approxvalidHS2}
\ee
\xxx{approxvalidHS2}
\yyy{p. DHS-18, boxes 2-6}
then
\be
\la\Delta \psi_{\cH_2}(\Delta t)|\Delta \psi_{\cH_2}(\Delta t)\ra^\frac{1}{2}=
\left[|c_1|^2E_1^2 + |c_2|^2E_2^2\right]^\frac{1}{2}\frac{\Delta t}{\hbar},
\ee
so the norm of the change in the state vector during the minimum possible time interval $\Delta t_{min}$\/ (assumed, as discussed above, to satisfy
(\ref{approxvalidHS2})) is greater than or equal to the limit (\ref{NLHS2}) deduced from the RQSL:
\be
\la\Delta \psi_{\cH_2}(\Delta t_{min})|\Delta \psi_{\cH_2}(\Delta t_{min})\ra^\frac{1}{2} \ge {\rm NormLim}_{{\rm RQSL} - \cH_2}
\ee

\subsection{Ideal measurement} \label{SubsectionIdealMeasurement}

We consider the simplest ideal measurement\ 
model, perhaps more appropriately described as  a model of a ``detector'' rather than a ``measuring device.''  An observer (a human or an apparatus) starts in a ready state $|\cO_1\ra$\/ and measures a two-state system
that starts and remains in its initial state.
If the  observer finds that the measured system is in state 
$|\cS_1\ra$\/ the observer  remains in the ready state; if she finds the measured system to be in state  $|\cS_2\ra$\/ the observer transitions to a state 
$|\cO_2\ra$\/.

In detail: The Hilbert space $\cH_D$\/ is spanned by the four vectors $|\cS_i\ra \otimes |\cO_j\ra,$\/ $i,j=1,2$\/, where
\be 
\la \cS_i|\cS_j\ra=\la \cO_i|\cO_j\ra=\delta_{i,j}, \hspace*{1cm} i,j=1,2. \label{HDstates}
\ee
\xxx{HDstates}
\yyy{p. DHS-20, first line}
The Hamiltonian is
\be
\wh{H}_{\cH_D}=\wh{P}^{\cS_2} \otimes \wh{h}^\cO \label{HamiltonianHD}
\ee
\xxx{HamiltonianHD}
\yyy{p. DHS-20, box 2}
where 
\be
\wh{P}^{\cS_2}=|\cS_2\ra\la \cS_2| \label{projectorS2}
\ee
\xxx{projectorS2}
\yyy{p. DHS-20, box 3}
and
\be
\wh{h}^\cO=i\kappa\left(|\cO_2\ra\la \cO_1| - |\cO_1\ra\la \cO_2|\right) \label{hO}
\ee
\xxx{hO}
\yyy{p. DHS-20, box 4}
The state vector describing the observer and measured system at the initial time $t=0$\/ has the observer in the ready state and uncorrelated with the measured system:
\be
|\psi_{\cH_D}(0)\ra=\left(c_1|\cS_1\ra + c_2|\cS_2\ra\right)\otimes|\cO_1\ra,\hspace*{1cm}|c_1|^2+|c_2|^2=1. \label{psiHD0}
\ee
\xxx{psiHD0}
\yyy{p. DHS-20, box 1}
Using (\ref{DeltaHdef}), (\ref{NLRQSLdef}) and (\ref{HDstates})-(\ref{psiHD0}), we find the 
lower limit on the norm of the change in the state vector from $t=0$\/ to $t=\Delta t_{min}$\/ to be
\be
{\rm NormLim}_{{\rm RQSL} - \cH_D} = \frac{|c_2|\kappa \Delta t_{min}}{\hbar}. \label{NLHSD}
\ee
\xxx{NLHSD}
\yyy{p. DHS-22, boxes 1,3}

The exact state vector for $t \ge 0$\/ is, from (\ref{HamiltonianHD})-(\ref{psiHD0}) and the Schr\"{o}dinger equation,\footnote{Note from (\ref{psitHD}) that at time $t_{meas}=\pi \hbar/(2\kappa)$\/ the ideal measurement has completed, in that the observer and observed-system states have become perfectly correlated.  This confirms that the choice of Hamiltonian (\ref{HamiltonianHD}) and initial state (\ref{psiHD0}) indeed models an ideal measurement. Of course $t_{meas}$\/ is much longer than $\Delta t_{min}$\/; 
see (\ref{DeltatHD}), (\ref{approxvalidHD}).}
\be
|\psi_{\cH_D}(t)\ra=c_1|\cS_1\ra|\cO_1\ra+c_2|\cS_2\ra\left[\cos\left(\frac{\kappa t}{\hbar}\right)|\cO_1\ra+\sin\left(\frac{\kappa t}{\hbar}\right)|\cO_2\ra\right] \label{psitHD}
\ee
\xxx{psitHD}
\yyy{p. DHS-26, box 1}
From (\ref{psitHD}) we see that the time scale characterizing changes in $|\psi_{\cH_D}(t)\ra$\/ is 
\be
\Delta t_{\cH_D}=\frac{2\pi \hbar}{|\kappa|}. \label{DeltatHD}
\ee
\xxx{DeltatHD}
We can arrive at the same conclusion by noting that the distinct eigenvalues of $\wh{H}_{\cH_D}$\/ are $0, \kappa$\/  and $-\kappa$\/, so
\be
\Delta t_{\cH_D}=2\pi \hbar/\max(|0|,|\kappa|,|-\kappa|). \label{DeltatHD2}
\ee
\xxx{DeltatHD2}
\yyy{p. DHS-30, last 4 boxes}
From (\ref{psitodeltapsi}), (\ref{HDstates}), (\ref{psiHD0}) and (\ref{psitHD}),
\be
\la\Delta \psi_{\cH_D}(t)|\Delta \psi_{\cH_D}(t)\ra^\frac{1}{2}=|c_2|\left[2-2\cos\left(\frac{\kappa t}{\hbar}\right)\right]^\frac{1}{2}.\label{exactnormHD}
\ee
\xxx{exactnormHD}
\yyy{p. DHS-27, box 4}
For $\Delta t$\/ satisfying
\be
\Delta t \ll \Delta t_{\cH_D} \label{approxvalidHD}
\ee
\xxx{approxvalidHD}
the norm of the change in the state vector, (\ref{exactnormHD}), becomes
\be
\la\Delta \psi_{\cH_D}(\Delta t)|\Delta \psi_{\cH_D}(\Delta t)\ra^\frac{1}{2}=\frac{|c_2|\kappa \Delta t}{\hbar}. \label{DeltatnormHD}
\ee
\xxx{DeltatnormHD}
\yyy{p. DHS-27, last box}
Taking $\Delta t$\/ to be the minimum possible time increment $\Delta t_{min}$\/, we see from (\ref{NLHSD}) and (\ref{DeltatnormHD})  that in this case the norm of the change in the state vector is equal to the limit coming from the RQSL:
\be 
\la\Delta \psi_{\cH_D}(\Delta t_{min})|\Delta \psi_{\cH_D}(\Delta t_{min})\ra^\frac{1}{2}= {\rm NormLim}_{{\rm RQSL} - \cH_D}.\label{NLRQSLHDcomparison}
\ee
\xxx{NLRQSLHDcomparison}

%
%
%
%
%
%

\section{Discussion}



While the arguments and results in \cite{BHZ05,CalmetHsu21,BHZ06,Hsu12,Hsu17,Hsu21,Rubin21} and the present paper suggest 
that the usual continuous Hilbert space is not the the proper arena in which to do quantum physics they do not  make this case conclusively, 
let alone 
point to 
a mathematical structure to serve as a replacement. 
Nevertheless Hilbert-space discreteness is an idea that 
deserves continued investigation, 
particularly for its potential to
explain one of the most recondite conceptual problems in physics and philosophy of science, that of the meaning and nature of probability.  

This issue is particularly salient for probability in the Everett interpretation of quantum mechanics.  In that theory, all outcomes of quantum processes occur in parallel, and since the inception of the theory physicists have struggled to understand the sense in which probabilities can be associated with outcomes 
if everything that can happen does happen \cite{Everett57,DeWitt72,Graham73,Okhuwa93,Kent90,Deutsch99,Barnumetal00,Rae09,Saunders10,Papineau10,Wallace10,GreavesMyrvold10,Kent10,Albert10,Price10,Zurek10,Schack10,Vaidman11,SebensCarroll18}.  

For example, when the ideal measurement of Sec.\ref{SubsectionIdealMeasurement} is complete, at time
 $t=t_{meas}=\pi \hbar/(2\kappa)$\/,
the state vector  (\ref{psitHD}) has the form 
\be
|\psi_{\cH_{D}}(t_{meas})\ra=c_1|\cS_1\ra|\cO_1\ra+c_2|\cS_2\ra |\cO_2\ra. \label{psitHDmeas}
\ee
\xxx{psitHDmeas}
In the standard (single-outcome, non-Everett) 
interpretation of quantum mechanics commonly presented in textbooks, 
a post-measurement state such as (\ref{psitHDmeas}) 
is replaced stochastically by 
one of the states corresponding to the possible outcomes of the measurement, either
$|\cS_1\ra|\cO_1\ra$\/ 
(``system is in state 1, observer is in the ready state 1'') or  
$|\cS_2\ra |\cO_2\ra$\/
 (``system is in state 2, observer is in state 2''), 
with the respective probabilities for each of these outcomes occurring being the absolute-squared amplitudes $|c_1|^2$\/, $|c_2|^2$\/.  This is termed ``reduction of the state vector''\cite[p. 236]{Ballentine98}.

In the Everett interpretation, on the other hand, there is no reduction.  The two terms on the right-hand side  of (\ref{psitHDmeas}) represents two Everett worlds. In one of these worlds the  outcome $|\cS_1\ra|\cO_1\ra$\/  has occurred, in the other the outcome $|\cS_2\ra |\cO_2\ra$\/ has occurred.

What can it possibly mean to say that one outcome is more or less probable than the other when in fact {\em both}\/ have occurred?\footnote{It is precisely the existence of multiple outcomes that is critical in enabling  the Everett interpretation to evade Bell's theorem and the resulting conclusion that quantum mechanics is nonlocal; see \cite[Sec. 1.1]{Rubin11} and references therein.}  This question is referred to as the ``incoherence problem''\cite{Wallace12}. 

The incoherence problem presents itself in a particularly acute form if we consider a situation in which one or more of the 
 outcomes is a  world corresponding to a state of affairs that is assigned such small probability by the standard quantum interpretation that we would feel justified in neglecting even the possibility of its occurring. For example, such a world might be one in which large violations of the second law of thermodynamics occur, with  heat flowing from colder to warmer bodies, gases and liquids spontaneously unmixing, etc.  Such worlds are termed ``maverick worlds''\cite{DeWitt70}.   Of course our intuition about the nature of probability leads us to feel that we understand why  any outcome with exceedingly small probability can be neglected.\footnote{Kolmogorov: ``We apply the theory of probability to the actual world of experiments in the following manner\ldots If P($A$\/) is very small, one can be practically certain that when conditions [for a repeatable experiment] are realized only once, the event $A$\/ would not occur at all\cite[pp. 3-4]{Kolmogorov56}.''} But, in the absence of a meaning we can assign to probability small or large, we are unable to explain why maverick outcomes are not features of our day-to-day experience. 

Incorporating Hilbert-space discreteness into Everett quantum mechanics provides  
a resolution to the problem of maverick worlds.  It implies that there is a minimum norm to state vectors, and one can then argue that maverick worlds, having norms below this minimum, are simply absent from any superposition of  outcomes\cite{BHZ06}. 

This might seem to still leave open the issue of how to assign meaning to probabilities {\em larger} than, say, those involving violation of the second law of thermodynamics---i.e., the probabilities encountered in the vast majority of physical situations.  But in fact, preclusion of state-vector components with norms below a certain minimum holds the potential of giving meaning to 
these probabilities as well.
%
%
%
For those probabilistic phenomena that are not precluded but that in fact may or may not be observed to occur, probability may be identified with  
subjective experience of a particular kind.  In the words of  de Finetti, one of the most prominent proponents of this subjective interpretation of  probability: ``What do we mean when we say, in ordinary language, that an event is more or less probable? We mean that we would be more or less surprised to learn that it has not happened\cite[p. 174]{deFinetti89}.''   Furthermore, 
there is a substantial body of {\em experimental}\/ evidence (see \cite[Sec. 6.4]{Rubin21} and references therein)  that the subjective experience of probability  is at its base a result of biological evolution\cite{Koopmans40}, \cite[Sec. 6.1]{Rubin21}. And it can be argued  that the  biological evolution, over generation upon generation, of subjective-probabilistic expectations fit to maverick worlds will be of sufficiently small norm to be precluded, leaving only organisms with subjective expectations matching our familiar notions of probability and statistics.
In a nutshell, ``preclusion explains evolution which then explains subjective probability\cite[p. 21]{Rubin21}.''
A proof-of-concept quantum-mechanical model of this process is presented in \cite[Secs. 6.2, 6.3]{Rubin21}. 

In addition to thus accounting for our experiences of probabilistic phenomena within the completely-deterministic Everettian framework, Everett quantum mechanics with preclusion of states below a minimum norm  
addresses  longstanding interpretational problems of probability {\em per se}\/\cite{SEPprobability}.  Along with purely-subjective theories of probability, this theory has the virtue of leaving no ambiguity as to what is meant by probability; namely, the subjective reactions of organisms to probabilistic  situations.   At the same time, the objective feature of preclusion 
allows it to provide 
{\em explanations}\/ 
for physical phenomena in general and for biological evolution in particular, something which purely subjective theories are incapable of doing.  An explanation should at the very least be 
``a statement of what is there in reality, and how it behaves and how that accounts for the explicanda\cite{Deutsch16}.'' So, the  explanation of  objectively-existing phenomena (including subjective probability\footnote{Subjective reactions are themselves objectively-existing phenomena.  ``The subjective opinion, as something known by the individual under consideration is, at least in this sense, something objective and can be a reasonable subject  of a rigorous study''\cite[p. 5]{deFinetti17}.}) must itself be something that exists objectively (in this theory, preclusion and the norms of quantum states).

For a detailed exposition of probability in the version of Everett quantum theory with preclusion outlined here, see \cite[Secs. 6.1, 7, 8]{Rubin21}. Other approaches to solving the Everett probability problem via discrete Hilbert space are presented in 
\cite{BHZ06,Hsu12, Hsu17, Hsu21}.  
Much more work needs to be done to place minimum Hilbert space norms on a concrete foundation, but the completion of such a program should prove highly consequential.



\section*{Acknowledgments}

I would like to thank Jianbin Mao, Jacob A. Rubin and Allen J. Tino for helpful discussions.

\end{document}